\documentclass[twocolumn]{aastex63}

\usepackage[T1]{fontenc}
\usepackage{ae,aecompl}
\usepackage{graphicx}
\usepackage{amsmath}
\usepackage{amssymb}
\usepackage{supertabular}
\usepackage{color}
\usepackage{longtable}
\usepackage{booktabs}

\shorttitle{Magnetar-driven X-ray plateaus}
\shortauthors{Hou et al.}

\begin{document}

\title{Evidence of X-ray plateaus driven by the magnetar spindown winds in gamma-ray burst afterglows}

\author{Shu-Jin Hou}
\affiliation{Department of Physics and Electronic Engineering, Nanyang Normal University, Nanyang, Henan 473061, China \\houshujingrb@163.com}
\affiliation{State Key Laboratory of Nuclear Physics and Technology, School of Physics, Peking University, Beijing 100871, China \\r.x.xu@pku.edu.cn}

\author{Shuang Du}
\affiliation{College of Mathematics and Physics, Wenzhou University, Wenzhou, Zhejiang 325035, China \\dushuang@pku.edu.cn}
\affiliation{State Key Laboratory of Nuclear Physics and Technology, School of Physics, Peking University, Beijing 100871, China \\r.x.xu@pku.edu.cn}

\author[0000-0001-8678-6291]{Tong Liu}
\affiliation{Department of Astronomy, Xiamen University, Xiamen, Fujian 361005, China \\tongliu@xmu.edu.cn}

\author{Hui-Jun Mu}
\affiliation{International Laboratory for Quantum Functional Materials of Henan and School of Physics and Microelectronics, Zhengzhou University, Zhengzhou, Henan 450001, China}

\author{Ren-Xin Xu}
\affiliation{State Key Laboratory of Nuclear Physics and Technology, School of Physics, Peking University, Beijing 100871, China \\r.x.xu@pku.edu.cn}
\affiliation{Kavli Institute for Astronomy and Astrophysics, Peking University, Beijing 100871, China}

\begin{abstract}
The central engine of gamma-ray bursts (GRBs) remains an open and forefront topic in the era of multimessenger astrophysics. The X-ray plateaus appear in some GRB afterglows, which are widely considered to originate from the spindown of magnetars. According to the stable magnetar scenario of GRBs, an X-ray plateau and a decay phase as $\sim t^{-2}$ should appear in X-ray afterglows. Meanwhile, the ``normal'' X-ray afterglow is produced by the external shock from GRB fireball. We analyze the Neil Gehrels \emph{Swift} GRB data, then find three gold samples, which have an X-ray plateau and a decay phase as $\sim t^{-2}$ superimposed on the jet-driven normal component. Based on these features of the lightcurves, we argue that the magnetars should be the central engines of these three GRBs. Future joint multimessenger observations might further test this possibility, then which can be beneficial to constrain GRB physics.
\end{abstract}


\section{Introduction}\label{sec1}

The central engine of gamma-ray bursts (GRBs) remains a mystery. The millisecond magnetars \citep[e.g.,][]{1992Natur.357..472U,1992ApJ...392L...9D,1998A&A...333L..87D,1998PhRvL..81.4301D,2001ApJ...552L..35Z,2007MNRAS.380.1541B,2011MNRAS.413.2031M,2020ApJ...901...75D} and black hole (BH) hyperaccretion \citep[e.g.,][]{Popham1999,Narayan2001,Kohri2005,Gu2006,Liu2007,2013ApJ...766...31K,Hou2014} are two main candidates of GRB central engines. For the recent reviews see \citet{Liu2017} and \citet{2018pgrb.book.....Z}. GRB prompt emission is generated by the internal shock in the ultra-relativistic jets, and the following afterglow is produced by the external shock \citep{1994ApJ...430L..93R,1998ApJ...497L..17S}. The lightcurves of prompt emission are usually irregular, but the afterglows generally have five components \citep[e.g.,][]{2006ApJ...642..354Z}.

The plateaus (shallow decays) are often seen in X-ray afterglow, which are usually thought to be caused by energy injected into the jets. The energy may be extracted from rotating magnetars or fallback accretions \citep[e.g.,][]{1998A&A...333L..87D,2001ApJ...552L..35Z,Liu2017,2019ApJ...886...87D,Huang2021}. Meanwhile, the superimposed normal decay is dissipated through the interaction between the jet and circumburst medium \citep[e.g.,][]{1998ApJ...497L..17S}. Their typical decay index is predicted to be $\sim 1.2$, but the observations are in the range of $0-1.5$ under the different circumstances and conditions \citep[e.g.,][]{1998ApJ...497L..17S,2006ApJ...642..354Z}. However, the spindown winds might not be injected into GRB jets, but dissipated behind GRB jets to power X-ray plateaus in the GRB magnetar model \citep{2020ApJ...901...75D}. In this scenario, the indices of the decays after plateaus are $\geq 2$ \citep{2001ApJ...552L..35Z}.

The magnetar could be born in the center of a massive collapsar or a neutron star (NS) binary merger \citep[e.g.,][]{2011PhRvD..83d4014G,2012LRR....15....8F,2021ApJ...908..106L}. For the hypermassive magnetar, the very short life leads that the spindown winds can not accumulate enough energy to produce observable features \citep{2000A&A...360..171R}. For the supramassive magnetar case, as discussed in \citet{2020ApJ...901...75D}, it (with life time $\gtrsim 100~\rm s$) has enough time to power the energetic spindown winds to generate the X-ray plateau followed by a steeper decay (with decay index $>3$). This phenomenon is called internal X-ray plateaus, which is understood as the collapse of a supramassive magnetar into a BH after the magnetar spindown \citep[e.g.,][]{2007ApJ...665..599T,2010MNRAS.409..531R,2017ApJ...849..119C,2018ApJ...854..104H}. In the stable magnetar case, the index of the decay following plateau is $\sim 2$ \citep{2001ApJ...552L..35Z}. It is interesting that the X-ray transient CDF-S XT2 with an X-ray plateau and a decay component as $\sim t^{-2}$ is well explained by the model of magnetar spindown winds under the stable magnetar scenario \citep{2019Natur...568...198X}.

We consider that the types of the GRB central engines can be judged by the similar observational features. Focus on the model of magnetar spindown winds, we propose the following criteria. First, there should exist an X-ray plateau and a decay component as $\sim t^{-2}$ in X-ray afterglows. However, this rule is not enough to prove a magnetar in the center of GRBs, because X-ray plateaus can be explained by involving the off-axis or precessing jets \citep[e.g.,][]{2020MNRAS.492.2847B,2020ApJ...893...88O,Huang2021} and the classical energy injections \citep[e.g.,][]{1998A&A...333L..87D,2001ApJ...552L..35Z}. Second, an additional X-ray radiation with decay index $\sim 1.2$ originating in the external shock should be superimposed with the plateau and a decay phase \citep[e.g.,][]{1998ApJ...497L..17S,2006ApJ...642..354Z}. This additional component indicates an X-ray plateau with a decay phase as $\sim t^{-2}$ is not produced by the external shock but the magnetar itself.

By using the above rules on the observations, the coincident sources could be satisfied with the model of magnetar spindown winds and further verify the existence of magnetars. The remaining part of this paper is organized as follows. We present our model in Section \ref{sec2}. The study of three cases is shown in Section 3. Section 4 is the conclusions and a brief discussion.

\section{Superpositions between GRB jets and magnetar spindown winds}\label{sec2}

\begin{figure}
\centering
\includegraphics[width=0.47\textwidth]{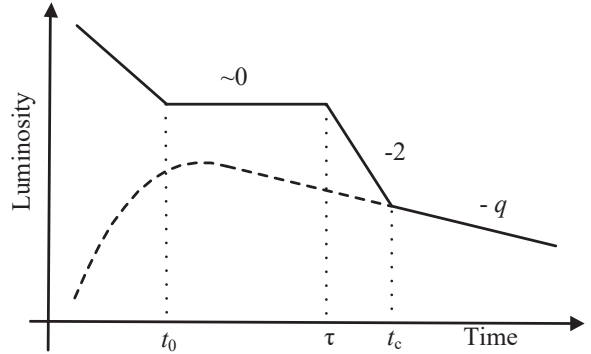}
\caption{Schematic diagram of X-ray lightcurve contributed by magnetar spindown winds and GRB jets. There is a ``tail'' of prompt emission before $t_{0}$, and the X-ray emission turns to be dominated by the spindown wind emission until $t>t_{\rm c}$. After $t_{\rm c}$, the X-ray emission is dominated by GRB jets. According to the standard external shock model, the typical value of $q$ is $\sim 1.2$.}
\label{fig1}
\end{figure}

For a GRB originated from a stable magnetar, as proposed in \cite{2020ApJ...901...75D}, its X-ray afterglow should be composed by the X-ray radiation from the winds driven by the magnetar spindown and magnetar-driven jets.

The evolution of the spindown luminosity of a magnetar can be expressed as
\begin{eqnarray}\label{1}
L_{\rm SD}= \frac{8\pi^{4}B_{\rm eff}^{2}R_{\ast}^{6}}{3c^{3}P^{4}} ,
\end{eqnarray}
where $B_{\rm eff}$ is the effective magnetic field strength on the NS surface (includes all the contribution that deviates from the dipole magnetic field), $R_{\ast}$ is the equatorial radius, and $P$ is the NS period \citep{2001ApJ...552L..35Z}.

If we consider that $B_{\rm eff}$ is a constant, the evolution of the X-ray luminosity $L_{\rm W}$ produced by the spindown winds is
\begin{eqnarray}\label{2}
L_{\rm W}= L_{\rm W,0}\left (1+ \frac{t}{\tau } \right )^{-2},
\end{eqnarray}
and the characteristic spindown time scale $\tau$ is
\begin{eqnarray}\label{3}
\tau =\frac{3c^{3}IP_{0}^{2}}{4\pi^{2}B_{\rm eff}^{2}R_{\ast}^{6}},
\end{eqnarray}
where $L_{\rm W,0}=\eta L_{\rm SD,0}$ is the initial X-ray luminosity of the spindown wind, $\eta$ is the efficiency of magnetic energy converting into X-ray emission, $L_{\rm SD,0}$ is the initial spindown luminosity, $I$ and $P_{0}$ are rotational inertia and initial period of the magnetar, and $t$ is the time from the burst, respectively. There is an X-ray plateau at $t<\tau$. When $t$ is much greater than $\tau$, there is a decay component as $\sim t^{-2}$.

\begin{figure}
\centering
\includegraphics[width=0.52\textwidth]{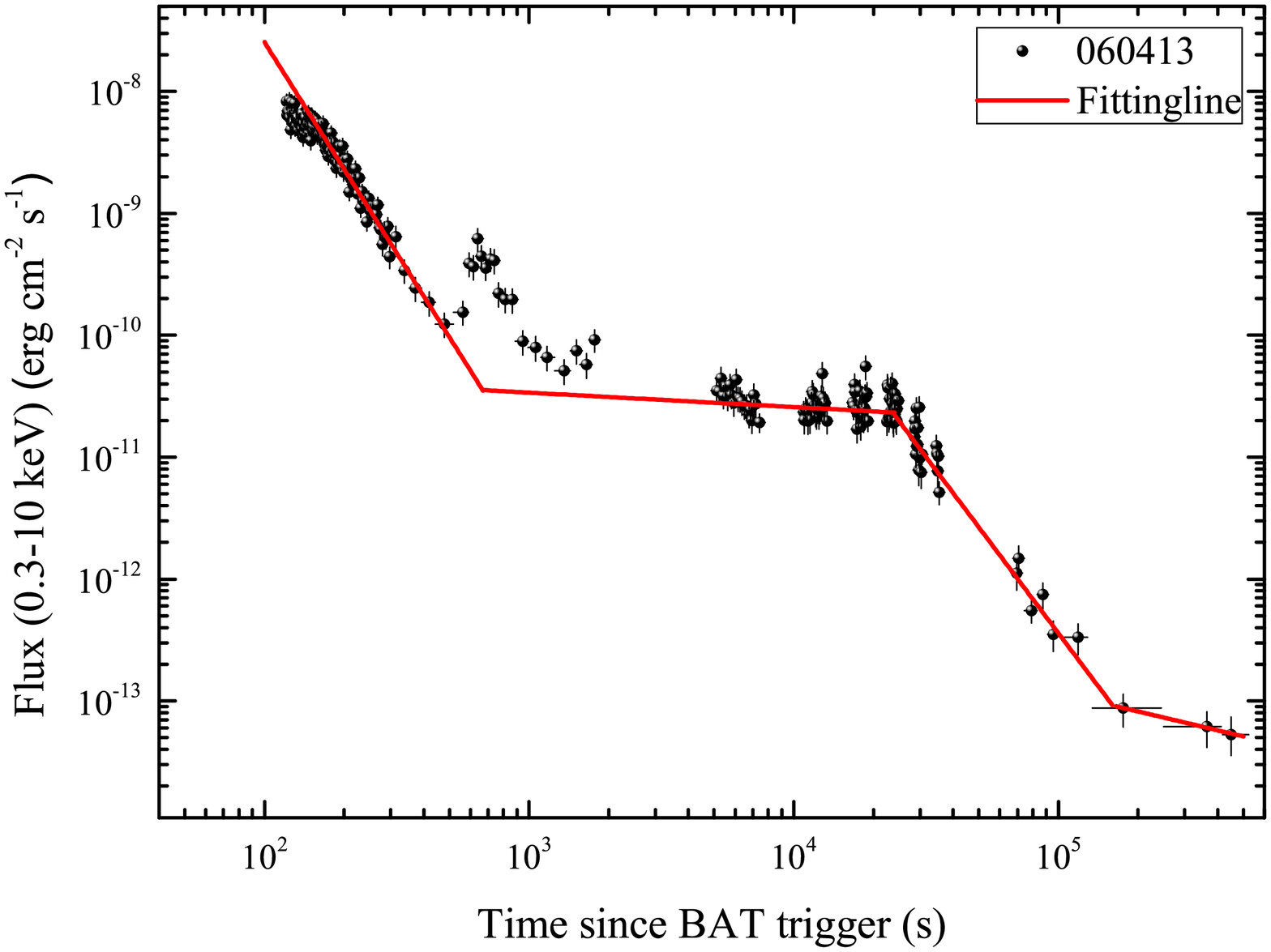}
\includegraphics[width=0.52\textwidth]{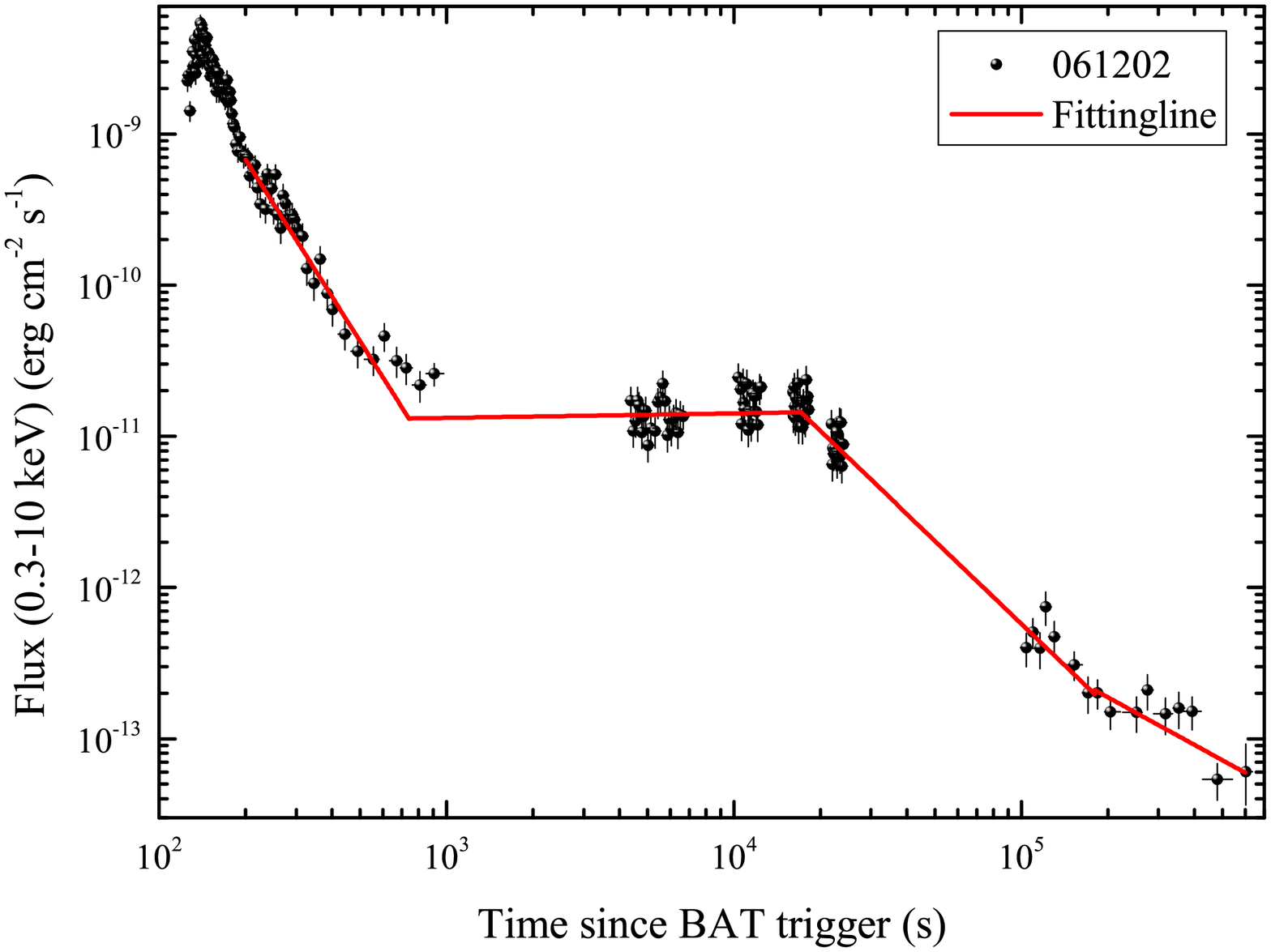}
\includegraphics[width=0.52\textwidth]{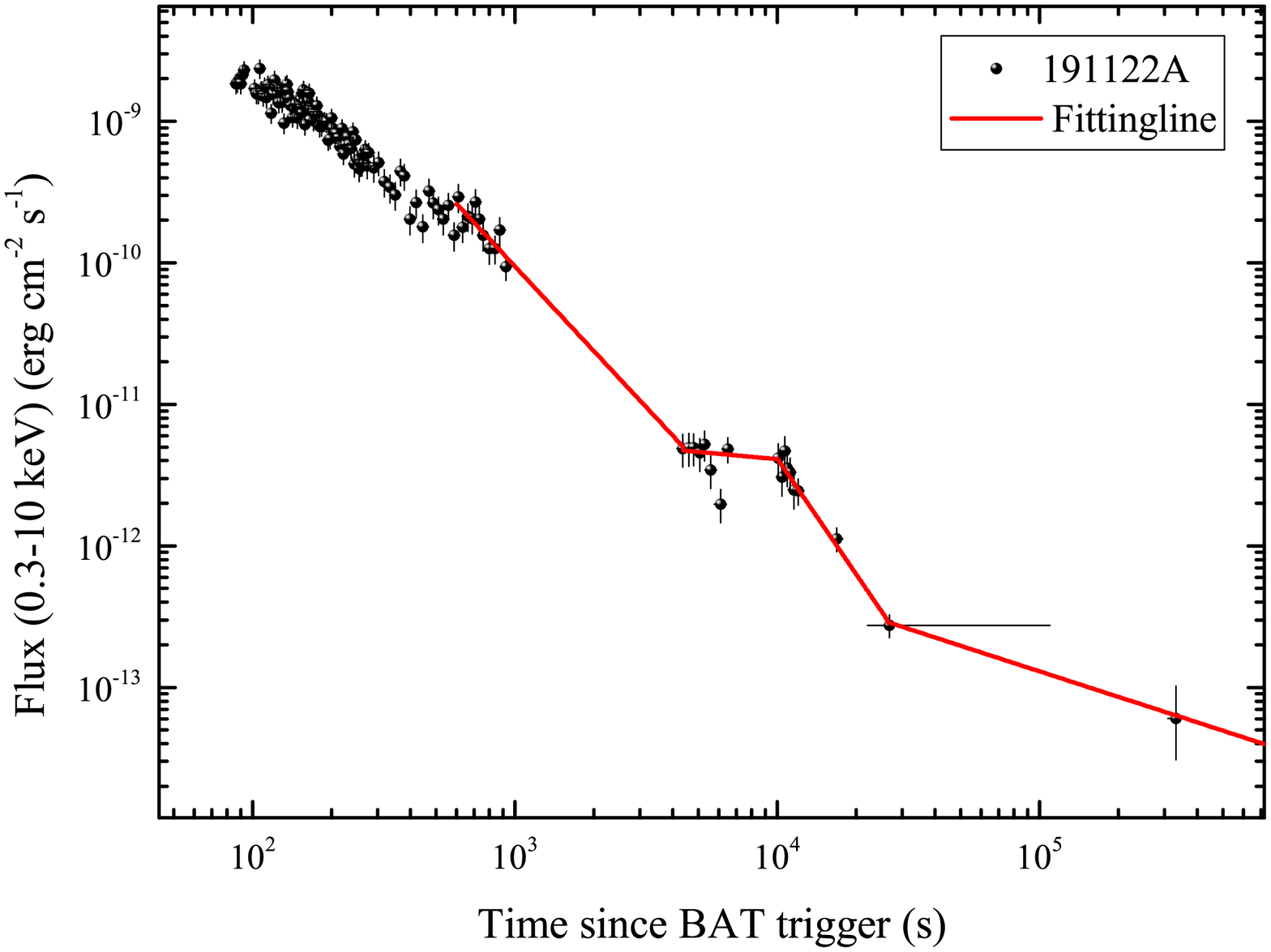}
\caption{X-ray lightcurves of GRBs 060413, 061202, and 191122A in 0.3$-$10 keV band. Red lines denote the best fitting lines with the multi-powerlaw function.}
\label{fig2}
\end{figure}

The luminosity of the X-ray emission from the jets, which is called the external shock model, can be empirically given by
\begin{eqnarray}\label{4}
L_{\rm J}=L_{\rm J,0}t^{-q},
\end{eqnarray}
where $L_{\rm J,0}$ is the X-ray luminosity of the jets and $q$ is the decay index with the typical value $\sim 1.2$ as shown in Figure \ref{fig1}.

Since the decay of the X-ray emission from GRB jets is slower than that from spindown winds after $\tau$ by comparison between Equations (\ref{2}) and (\ref{4}), there is a situation that $L_{\rm W}$ is always smaller than $L_{\rm J}$. There is also the possibility of another situation that the whole afterglow is dominated by the spindown winds. In those situations, the components from non-dominant contributions are hardly identified through the lightcurve. We are not interested in those situations and will not discuss here.

It is worth discussing the situation that the X-ray emission is alternately dominated by the spindown winds and GRB jets. In early stage of the X-ray afterglow, the ``tail'' of prompt emission and the emission from spindown winds may be very strong, so the X-ray emission from GRB jets may be masked. Here we only discuss that the early phases of X-ray afterglows are dominated by spindown winds and the later phases are dominated by GRB jets.

\begin{figure}
\centering
\includegraphics[width=0.52\textwidth]{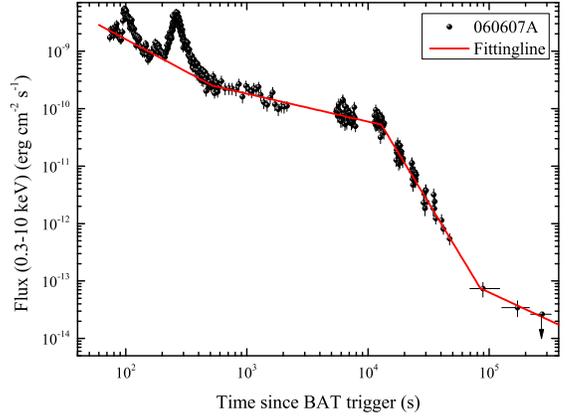}
\caption{X-ray lightcurves of GRB 060607A in 0.3$-$10 keV band. The slope after the plateau is 3.48. In this case, an NS may collapse into a BH at the later stage.}
\label{fig3}
\end{figure}

\begin{table*}
\begin{center}
\caption{Fitting results of X-ray lightcurves with multi-powerlaw functions.}
\begin{tabular}{ccccc}
\hline
\hline
GRBs & $\alpha_{1}$ (err) & $\alpha_{2}$ (err) & $\alpha_{3}$ (err) & $\alpha_{4}$ (err) \\
\hline
060413 & 3.46 (0.13) & 0.12 (0.05)  & 2.89 (0.14) & 0.52 (0.04) \\
061202 & 3.02 (0.17) & -0.03 (0.06)  & 1.83 (0.05) & 1.04 (0.29) \\
191122A & 1.99 (0.12) & 0.17 (0.22)  & 2.72 (0.14) & 0.60 \\
\hline
060607A & 1.13 (0.04) & 0.49 (0.02)  & 3.48 (0.11) & 0.98 (0.18)\\
\hline
\hline
\end{tabular}
\begin{minipage}{12cm}
\emph{Notes}: \\(1) $\alpha_{1}$ denotes the slope of the steep decay component (the ``tail'' of prompt emission); $\alpha_{2}$ and $\alpha_{3}$ represent the slopes of the plateau and the following decay component, respectively; and $\alpha_{4}$ corresponds the decay index of the component at later stage. \\(2) Due to a lack of late-time date, $\alpha_4$ is fixed for GRB 191122A. $\alpha_3$ for GRB 060607A is $\sim 3.48$, which indicates a collapsing NS progenitor and is not suitable for our sample.
\end{minipage}
\end{center}
\end{table*}

Based on the relation of $L_{\rm W}=L_{\rm J}$, we can obtain the solutions as
\begin{eqnarray}\label{5}
\begin{cases}
t_{\rm c}=\left ( \frac{L_{\rm W,0}}{L_{\rm J,0}} \right )^{\frac{1}{-q}} & \text{ if } t\ll \tau \\
t_{\rm c}=\left ( \tau^{2}\frac{L_{\rm W,0}}{L_{\rm J,0}} \right )^{\frac{1}{2-q}} & \text{ if } t\gg \tau
\end{cases}.
\end{eqnarray}

Therefore, if there is a ``tail'' of prompt emission before $t_{0}$, the first solution $t_{\rm c}$ is not visible in the lightcurve, then the X-ray emission turns to be dominated by the spindown wind emission until $t>t_{\rm c}$.  After $t_{\rm c}$, the X-ray emission is dominated by the jets. The corresponding lightcurve of three typical example is shown in Figure \ref{fig1}.

It needs to be emphasized that the X-ray plateaus can be explained by the off-axis or precessing jets with whatever type of central engines. However, these models cannot explain the decay index changes at later stage of X-ray afterglows. The inflexion means that there are two different components, which is predicted by the model of the magnetar spindown winds. The X-ray plateaus with the following decay segments as $\sim t^{-2}$  is powered by a magneter spindown winds. The X-ray afterglows following the spindown wind segment is from the standard external shock of the jets.

\section{Samples}

According to the model discussed above, we select candidates according to the following criteria: (i) a plateau should exist in the X-ray emission lightcurve; (ii) after the plateau, there is a steeper decay with index $\sim 2$; (iii) at the later stage of the X-ray lightcurve, there is another decay component with index $\sim 1.2$. This component is the key criteria of our sample.

According to the above criteria, we found three gold samples in the Neil Gehrels \emph{Swift} GRB data, i.e., GRBs 060413, 061202, and 191122A. All of them belong to the long-duration GRBs. The data are from the UK \emph{Swift} Science Data Center at the University of Leicester \citep{2007A&A...469..379E,2009MNRAS.397.1177E}. The afterglows of these bursts all contain multiple components. There are significant flares in GRBs 060413 and 061202. At late times, the decay index of afterglow can be constrained even though the data are sparse.

We employ the multiple broken power-law functions to fit their X-ray lightcurves. The fitting results of four indices $\alpha_1$, $\alpha_2$, $\alpha_3$, and $\alpha_4$ are listed in Table 1. The plateau component are all flat with the decay indices less than $\sim 0.2$. After the plateau, their decay indices are 2.89, 1.83, and 2.72, respectively. These values are basically consistent with the slope of the magnetar spindown process. The indices of the last components are 0.52, 1.04, and 0.60, respectively. Under the different circumstances and conditions, the slope range of the lightcurve from the external shock model could be 0$-$1.5 \citep[e.g.,][]{1998ApJ...497L..17S,2006ApJ...642..354Z}, so we reasonably believe that these components come from GRB jets and further the central engines of these three GRBs should be magnetars.

In our samples, only the redshift of GRB 061202 is measured, i.e., $z=2.25$. The X-ray luminosity of plateau $L_{\rm W}$ can be expressed as
\begin{eqnarray}\label{6}
L_{\rm W}= \frac{4\pi D_{\rm L}^{2}F_{\rm W}}{1+z},
\end{eqnarray}
where $D_{\rm L}$ is the luminosity distance and $F_{\rm W}$ is the flux of plateau. By fitting, we get that $F_{\rm W}$ and $\tau$ are $\sim1.0 \times10^{-11}$ erg s$^{-1}$ cm$^{-2}$ and $\sim1.8\times10^{4}$ s, respectively. Then $L_{\rm W}$ can be estimated to be $\sim4.0 \times10^{48}$ erg s$^{-1}$. Assuming that the rotational inertia of the magnetar is $10^{45}$ g cm$^{2}$ and the radius of NS is 10 km \citep[e.g.,][]{Li2020}, we can obtain that the isotropic energy of plateau $E_{P}\sim F_{\rm W}\tau \simeq 2.4\times10^{52}$ ergs. According to Equations (\ref{1}) and (\ref{3}), the effective magnetic field strength on the NS surface $B_{\rm eff}$ and initial period $P_{0}$ are calculated, $\sim 5.0 \times10^{14}$ G and $\sim 1.1$ ms, respectively. These values are satisfied with the magnetar model.

\section{Conclusions and Discussion}

In this paper, we analyze the shapes of lightcurves produced by magnetars spindown winds, which have an X-ray plateau and a decay component as $\sim t^{-2}$. According to the fireball model of GRBs, the X-ray afterglow is produced by the external shock from jets at the same time. However, the ``tail'' of prompt emission and the emission from spindown winds may be powerful at the early stage of X-ray afterglow, so the X-ray emission from the jets may be masked. We only discussed that the early phases of X-ray afterglow are dominated by the spindown winds and the later phases are dominated by jets. So there is a state transition in the lightcurve, which is a change in slope at the later stage of afterglows. We emphasize that the presence of the jet component is very important, because it supports the idea that the X-ray plateau arises from a component with a different origin. By systematically analyzing the XRT lightcurves of GRBs detected by the Neil Gehrels \emph{Swift} observatory, we find three gold samples to be consistent with the model of the magnetar spindown winds. Taking GRB 061202 as an example, the $E_{P}$, $B_{\rm eff}$, and $P_{0}$ are estimated. They are all within the ranges of typical magnetar parameters. Since the features of the detectable MeV neutrinos and gravitational waves from (newborn) magnetars and BH hyperaccretion are distinguishable \citep[e.g.,][]{2016PhRvD..93l3004L,2019ApJ...878..142W,2020ApJ...889...73W}, future joint multimessenger observations might provide more evidences on the magnetar-driven GRBs.

The above discussion is based on a stable magnetar. If the magnetar is a supramassive one, the corresponding X-ray lightcurve is similar to that in Figure \ref{fig1}, but the spindown time $\tau$ should be changed to the break time $t_{\rm B}$ (corresponding to the collapse time of the magnetar), and the slope of the segment following the plateau should be steeper (with decay index $>3$). For example, the decay index following the plateau is 3.48 in GRB 060607A as shown in Figure \ref{fig3}. In some GRBs, the slopes even go up to $\sim 9$ \citep[e.g.,][]{2007ApJ...665..599T}. The internal X-ray plateaus is thought to go through a spindown process and then collapse into a BH \citep[e.g.,][]{2007ApJ...665..599T,2017ApJ...849..119C,2018ApJ...854..104H}.

A magnetar might be the central engine of X-ray transient CDF-S XT2 \citep[e.g.,][]{2019Natur...568...198X,2019ApJ...879L...7X}, this process only explains the spindown wind of the magnetar and could not see the slow decay which is powered by external shock. In our scenario, we consider that the components including the contributions of the jets and the magnetar spindown winds are the more reliable evidence for the existence of magnetars.

\section*{Acknowledgement}
We acknowledge the use of the public data from the Neil Gehrels \emph{Swift} data archive and the UK \emph{Swift} Science Data Center at the University of Leicester. This work is supported by the National Natural Science Foundation of China under grants U1938116, 11822304, 12173031, and 12103047, the National Key R\&D Program of China No. 2017YFA0402602, the National SKA Program of China No. 2020SKA0120100, the Natural Science Foundation of Henan Province of China under grant 212300410290, and China Postdoctoral Science Foundation under grant 2019TQ0288.

\end{document}